\documentclass[twocolumn]{jpsj2}
\usepackage{graphicx}
\usepackage{amsmath,amssymb}

\def\runauthor{Youichi {\sc Yanase}, M. Motizuki and Masao {\sc Ogata}}

\title{ 
Role of Spin-Orbit Coupling on the Spin Triplet Pairing in 
${\rm Na_{x}Co_{}O_{2}} \cdot y{\rm H}_{2}{\rm O}$ 
\\ 
I: $d$-vector under Zero Magnetic Field
} 

\author{Youichi {\sc Yanase}\footnote{E-mail:
yanase@hosi.phys.s.u-tokyo.ac.jp}, Masahito {\sc Mochizuki} 
and Masao {\sc Ogata}}

\inst{Department of Physics, University of Tokyo, Tokyo 113-0033}

\recdate{Today 2005}

\abst
{
 The $d$-vector in possibile spin triplet superconductor 
${\rm Na_{x}Co_{}O_{2}} \cdot y{\rm H}_{2}{\rm O}$ is microscopically 
investigated on the basis of the multi-orbital Hubbard model including 
the atomic spin-orbit coupling. 
 As a result of the perturbation theory, we obtain the stable 
spin triplet superconductivity where the $p$-wave and $f$-wave states 
can be stabilized. 
 If we neglect the spin-orbit coupling, superconducting state has 
$6$-fold ($3$-fold) degeneracy in the $p$-wave ($f$-wave) state. 
 This degeneracy is lifted by the spin-orbit coupling. 
 We determine the $d$-vector within the linearlized 
Dyson-Gorkov equation. 
 It is shown that the $d$-vector is always along the plane when the 
pairing symmetry is $p$-wave, while it depends on the parameters 
in case of the $f$-wave state. 
 The lifting of degeneracy is significant in the $p$-wave state 
while it is very small in the $f$-wave state. 
 This is because the first order term with respect to the spin-orbit 
coupling is effective in the former case, while it is ineffective 
in the latter case. 
 The consistency of these results with NMR and $\mu$SR measurements 
are discussed. 
}

\kword
{
${\rm Na_{x}Co_{}O_{2}} \cdot y{\rm H}_{2}{\rm O}$; 
spin triplet superconductivity; 
multi-orbital model;
$d$-vector
}

\begin{document}
\sloppy
\maketitle

\newcommand{\eli}{$\acute{{\rm E}}$liashberg }
\renewcommand{\k}{\mbox{\boldmath$k$}}
\newcommand{\q}{\mbox{\boldmath$q$}}
\newcommand{\Q}{\mbox{\boldmath$Q$}}
\newcommand{\kk}{\mbox{\boldmath$k'$}}
\newcommand{\e}{\varepsilon}
\newcommand{\ee}{\varepsilon^{'}}
\newcommand{\s}{{\mit{\it \Sigma}}}
\newcommand{\J}{\mbox{\boldmath$J$}}
\newcommand{\vv}{\mbox{\boldmath$v$}}
\newcommand{\Jh}{J_{{\rm H}}}
\newcommand{\LL}{\mbox{\boldmath$L$}}
\renewcommand{\SS}{\mbox{\boldmath$S$}}
\newcommand{\Tc}{$T_{\rm c}$ }
\newcommand{\Tcf}{$T_{\rm c}$}
\newcommand{\Co}{${\rm Na_{x}Co_{}O_{2}} \cdot y{\rm H}_{2}{\rm O}$ }
\newcommand{\Cof}{${\rm Na_{x}Co_{}O_{2}} \cdot y{\rm H}_{2}{\rm O}$}
\newcommand{\tgf}{$t_{\rm 2g}$-orbitals}
\newcommand{\tg}{$t_{\rm 2g}$-orbitals }
\newcommand{\av}{\mbox{\boldmath${\rm a}$} }
\newcommand{\bv}{\mbox{\boldmath${\rm b}$} }
\newcommand{\avf}{\mbox{\boldmath${\rm a}$}}
\newcommand{\bvf}{\mbox{\boldmath${\rm b}$}}
\newcommand{\egf}{$e_{\rm g}$-Fermi surface }
\newcommand{\egff}{$e_{\rm g}$-Fermi surface}
\newcommand{\agf}{$a_{\rm 1g}$-Fermi surface }
\newcommand{\agff}{$a_{\rm 1g}$-Fermi surface}

\section{Introduction}

 The unconventional superconductivity in strongly correlated 
systems has been one of the central issues in the condensed matter 
physics. For example, heavy fermion superconductors,~\cite{rf:steglich} 
high-\Tc cuprates,~\cite{rf:bednorz} 
organic superconductors~\cite{rf:kanoda} 
and Sr$_2$RuO$_4$~\cite{rf:maeno} are cited. 
 Recently, a new superconductor \Co was discovered by Takada 
{\it et al},~\cite{rf:takada} and the possibility of unconventional 
superconductivity has attracted huge interests.

 Immediately after the discovery of superconductivity in \Cof, 
the symmetry of superconductivity has been studied by 
many experimental measurements.~\cite{
rf:yoshimura,rf:kobayashi,rf:zheng,rf:ishida,rf:higemoto,
rf:uemura,rf:kanigel,rf:hdyang,rf:lorenz,
rf:oeschler,rf:sasaki,rf:yokoi,rf:sakuraireview} 
 While some controversial results exist,~\cite{rf:yokoi} 
most of them have indicated the non-$s$-wave superconductivity. 
 For example, the absence of coherence peak in NMR 
$1/T_1 T$~\cite{rf:zheng,rf:ishida} and the power low behaviors in 
$1/T_1 T$~\cite{rf:zheng,rf:ishida} and specific 
heat~\cite{rf:hdyang,rf:lorenz,rf:oeschler} are evidences for the  
anisotropic pairing. 
 Recently, a magnetic phase has been discovered in the neighborhood of 
superconducting phase.~\cite{rf:ihara,rf:sakuraimagnetic} 
 This observation clearly indicates the importance of electron 
correlation which generally leads to the non-$s$-wave superconductivity.

 These experimental indications have accelerated theoretical studies
on the superconductivity in \Cof. 
 In the first stage, the effect of frustration stimulated many interests
since \Co has layered structure constructed by the triangular lattice.  
 The RVB theory has been applied to the triangular 
lattice~\cite{rf:baskaran,rf:shastry,rf:lee,rf:ogata} and 
concluded the spin singlet $d$-wave superconductivity. 
 Some authors have pointed out the frustration of charge ordering 
at the electron filling $n=4/3$,~\cite{rf:lee} and 
the $f$-wave superconductivity due to the charge fluctuation 
has been discussed.~\cite{rf:Ytanaka,rf:motrunich}  
 Recently, the RVB theory has been applied to the multi-orbital model 
and concluded the spin triplet superconductivity.~\cite{rf:khaliullin}

 Another theoretical approach is based on the perturbation expansion 
from the weak coupling region which includes the 
perturbation theory,~\cite{rf:nisikawa} 
random phase approximation,~\cite{rf:yata}
FLEX approximation~\cite{rf:kuroki} and 
perturbative renormalization group.~\cite{rf:honerkamp}  
 Some authors have taken account of the Fermi surface of \Co partly, 
and concluded the $f$-wave superconductivity,~\cite{rf:kuroki} 
$g$-wave superconductivity,~\cite{rf:yata} 
$i$-wave superconductivity~\cite{rf:kurokii-wave} and 
nearly degeneracy between $d$- and $f$-wave 
superconductivities.~\cite{rf:nisikawa}

 Although these theories except for Refs.~27 and 29 have assumed 
single-orbital models, the conduction band in \Co has orbital 
degeneracy, as pointed 
out by Koshibae {\it et al.}~\cite{rf:koshibae} 
 The conduction band mainly consists of three $t_{\rm 2g}$-orbitals 
in Co ions which hybridize with O2p-orbitals. 
 Therefore, it is highly desired that the pairing symmetry 
in this material is examined on the basis of the multi-orbital model. 
 
 For this purpose, we have constructed a three-orbital Hubbard model 
which appropriately reproduces the electronic structure obtained 
in the LDA calculations,~\cite{rf:singh,rf:pickett,rf:arita} 
and we have analyzed it on the basis of the FLEX 
approximation~\cite{rf:mochizuki} and perturbation theory.~\cite{rf:yanase} 
 Then, we have obtained some notable results which are summarized as 
follows. (i) The spin triplet superconductivity is stable unless the 
Hund's rule coupling is very small. (ii) The $p$-wave state and 
$f$-wave state are nearly degenerate owing to the orbital degree of 
freedom. (iii) There is a nearly ferromagnetic spin correlation along 
the plane which stabilizes the spin triplet pairing. 
 This is consistent with recent neutron scattering measurements 
which have reported the anti-ferromagnetic order with stacking 
ferromagnetic plane.~\cite{rf:neutron} 
(iv) The vertex correction whose importance has been pointed out 
for Sr$_2$RuO$_4$~\cite{rf:yanasereview,rf:nomura} is not important 
in \Cof. 
(v) The two of three orbitals, 
namely $e'_{\rm g}$-doublet are essential for the superconductivity. 
 The orbital-dependent-superconductivity proposed for 
Sr$_2$RuO$_4$~\cite{rf:agterberg} is partly justified in \Cof. 
(vi) However, the orbital degeneracy in $e'_{\rm g}$-doublet is 
particularly important because the single-orbital approximation 
artificially suppresses the $p$-wave superconductivity. 
 This point is in sharp contrast to Sr$_2$RuO$_4$ 
where the single-orbital approximation is valid.~\cite{rf:nomura} 
 In this sense, \Co is a typical multi-orbital superconductor 
in $d$-electron system.

 One of the interesting subjects in the multi-orbital superconductor is 
the role of spin-orbit coupling. 
 This issue is particularly important in the spin triplet 
superconductivity which has an internal degree of freedom described 
by the $d$-vector.~\cite{rf:leggett,rf:sigrist} 
 If we neglect the spin-orbit coupling as in the 
previous studies,~\cite{rf:mochizuki,rf:yanase} the $6$-fold ($3$-fold) 
degeneracy remains in the $p$-wave ($f$-wave) state at $T=T_{\rm c}$.  
 Therefore, we have to take into account the spin-orbit coupling 
to determine the pairing state. 
 The goal of this paper is to clarify the role of spin-orbit 
coupling and microscopically determine the $d$-vector in \Cof. 

 The $d$-vector is particularly important to discuss the Knight shift 
measurement which has been a powerful method to determine the pairing
symmetry.~\cite{rf:tou-ishida} 
 This is because the magnetic susceptibility in spin triplet superconductor 
significantly depends on the direction of $d$-vector. 
 Although many experiments have been performed to determine 
the pairing symmetry in \Cof,~\cite{rf:yoshimura,
rf:kobayashi,rf:zheng,rf:ishida,rf:higemoto,rf:uemura,rf:kanigel,
rf:hdyang,rf:lorenz,rf:oeschler,rf:yokoi,rf:sasaki,rf:sakuraireview} 
there is any conclusive evidence neither for the spin triplet pairing 
nor for the spin singlet pairing.  
 We think that this is partly due to the lack of knowledges 
on the pairing state expected in the spin triplet superconductivity. 
 The results in this paper will provide a clear subject for 
a comparison between the theory and experiment.

 The role of spin-orbit coupling on the spin triplet superconductivity  
has been a longstanding problem since the discovery of heavy fermion 
superconductors.~\cite{rf:sigrist,rf:tou-ishida,rf:anderson,rf:machida,
rf:sauls,rf:joynt} However, the microscopic study has not been performed 
owing to the complicated electronic structure of heavy fermion systems. 
 The discussion about the pairing symmetry in UPt$_3$ still 
continues~\cite{rf:machida,rf:sauls,rf:joynt} partly because there is no 
microscopic research on the anisotropy of $d$-vector. 
 Recently, we have developed a microscopic theory on the $d$-vector 
and applied to Sr$_2$RuO$_4$.~\cite{rf:yanaseRuSO} 
 The present study on \Co provides a contrasting example and 
we expect that these studies on the $d$-electron systems will lead to 
a systematic understanding including the $f$-electron systems.

 This paper is organized as follows. 
 In \S2, we introduce the three-orbital Hubbard model including the 
atomic spin-orbit coupling and derive the two-orbital Hubbard model like
in the previous study.~\cite{rf:yanase} 
 The pairing symmetry allowed in this model is classified in \S3. 
 The linearized Dyson-Gorkov equation in the multi-orbital model 
including the spin-orbit coupling is developed in \S4.1. 
 In \S4.2, the pairing state is determined on the basis of 
the second order perturbation theory. We show that the role of 
spin-orbit coupling is quite different between the $p$-wave 
superconductivity and $f$-wave one. 
 This difference is illuminated by showing the splitting of \Tc 
in \S4.3. We show that the $d$-vector in the $p$-wave state is 
strongly fixed against the magnetic field, while that in the 
$f$-wave state is fixed weakly. 
 In \S4.4, we discuss the cross-over from the weak spin-orbit coupling 
region $\lambda \ll W$ relevant for \Co to the strong spin-orbit
coupling region $\lambda \gg W$. 
 Although the latter is unrealistic for \Cof, 
this analysis will be useful for a unified understanding including 
the heavy fermion superconductors. 
 In \S5, we summarize the comparisons between \Co and Sr$_2$RuO$_4$. 
 Some discussions are given in the last section \S6.

\section{Spin-Orbit Coupling in Multi-Orbital Hubbard Model}

 First of all, we introduce a three-orbital Hubbard model including 
the spin-orbit coupling from which a two-orbital model are derived later.  
 We consider a two-dimensional model 
which represents Co ions on the triangular lattice. 
 We have constructed a tight-binding model for Co \tg 
which reproduces the results
of LDA calculations.~\cite{rf:mochizuki} 
 By adding the spin-orbit coupling term and Coulomb interaction term, 
the three-orbital Hubbard model is obtained as, 
\begin{eqnarray}
&& \hspace{-5mm} H_{3} = H_{0}+H_{\rm LS}+H_{{\rm I}}, 
\\
&& \hspace{-5mm} H_{0} = \sum_{i,j,s} \sum_{a,b} t_{a,b,i,j} 
c_{i,a,s}^{\dag} c_{j,b,s},  
\\
&& \hspace{-5mm} H_{\rm LS}=2 \lambda \sum_{i} \LL_{i}  \SS_{i},
\\
&& \hspace{-5mm} H_{{\rm I}} = 
U \sum_{i} \sum_{a} n_{i,a,\uparrow} n_{i,a,\downarrow} 
+ U' \sum_{i} \sum_{a>b} n_{i,a} n_{i,b} 
\nonumber \\
&& \hspace{0mm}
- \Jh \sum_{i} \sum_{a>b} (2 \SS_{i,a} \SS_{i,b} 
+ \frac{1}{2} n_{i,a} n_{i,b})
\nonumber \\
&& \hspace{0mm}
+ J \sum_{i} \sum_{a \neq b}  
c_{i,a,\downarrow}^{\dag} 
c_{i,a,\uparrow}^{\dag} 
c_{i,b,\uparrow} 
c_{i,b,\downarrow}. 
\label{eq:multi-orbital-model}
\end{eqnarray}
 Here, the indices $i$ and $j$ denote the sites in the real space and 
indices $a$ and $b$ denote the orbitals. 
 We assign the $d_{\rm xy}$-, $d_{\rm yz}$- and $d_{\rm xz}$-orbitals 
to $a=1$, $a=2$ and $a=3$, respectively. 

 The first term $H_{0}$ is a tight-binding Hamiltonian where 
$t_{a,b,i,j}$ are determined according to the symmetry of 
orbitals and lattice. 
 The dispersion relation expected in the LDA calculation is reproduced 
by assuming nine hopping parameters from $t_1$ to $t_9$ as well as 
the crystal field splitting $e_{\rm c}$ which arises from 
the distortion of octahedron. 
 For instance, we assume $t_{1,1,i,i \pm \avf}=t_{1}$, 
$t_{1,1,i,i \pm \bvf}=t_{2}$, $t_{2,3,i,i \pm \avf}=t_{3}$, 
$t_{1,1,i,i \pm (\avf-\bvf)}=t_{4}$, $t_{1,1,i,i \pm 2\avf}=t_{5}$, 
$t_{2,3,i,i \pm 2\avf}=t_{6}$, $t_{2,3,i,i \pm (\avf-\bvf)}=t_{7}$, 
$t_{1,3,i,i \pm (\avf-\bvf)}=t_{8}$ and 
$t_{1,2,i,i \pm (\avf-\bvf)}=t_{9}$.  
 We denote the basis of triangular lattice as \avf=$(\sqrt3/2,-1/2)$ 
and \bvf=$(0,1)$ and we choose the lattice constant as a unit length. 
 The other hopping matrix elements $t_{a,b,i,j}$ are obtained 
by the symmetry operation. 
 Since $t_{3}$ is largest among $t_{n}$, we choose the unit of energy as 
$t_{3}=1$ throughout this paper. 
 We describe $H_{0}$ in the matrix representation as, 
\begin{eqnarray}
 \label{eq:three-band-model-kinetic}
 H_0 = \sum_{\k,s} \hat{c}_{\k,s}^{\dag} \hat{H}(\k) \hat{c}_{\k,s}, 
\end{eqnarray} 
where $\hat{c}_{\k,s}^{\dag}=
(c_{\k,1,s}^{\dag},c_{\k,2,s}^{\dag},c_{\k,3,s}^{\dag})$ is a vector 
representation of Fourier transformed creation operators with spin $s$. 
 The matrix element of $\hat{H}(\k)$ has been given in Refs.~37 and 38.

 In order to clarify the nature of superconductivity in this model, 
it is useful to introduce a non-degenerate $a_{\rm 1g}$-orbital 
and doubly-degenerate $e'_{\rm g}$-orbitals.  
 They are defined as, 
\begin{eqnarray} 
\label{eq:a1g}
&& \hspace{-5mm} 
|a_{\rm 1g}> = \frac{1}{\sqrt{3}}(|{\rm xy}>+|{\rm xz}>+|{\rm yz}>), 
\\
\label{eq:e1}
&& \hspace{-5mm} 
|e_{\rm g}, 1> = \frac{1}{\sqrt{2}}(|{\rm xz}>-|{\rm xy}>),
\\
\label{eq:e2}
&& \hspace{-5mm} 
|e_{\rm g}, 2> = \frac{1}{\sqrt{6}}(2|{\rm yz}>-|{\rm xz}>-|{\rm xy}>). 
\end{eqnarray}
 The orbital-dependent-superconductivity in \Co occurs in 
this basis as shown in Refs.~37 and 38. 
 We choose a basis of $e'_{\rm g}$-orbitals different from Ref.~38
in order to simplify the following notations. 
 By choosing the basis wave function as 
eqs.~(\ref{eq:a1g}-\ref{eq:e2}), 
the tight-binding Hamiltonian is transformed to be, 
\begin{eqnarray}
 \label{eq:unitary-local}
H_0 = \sum_{\k,s} \hat{d}_{\k,s}^{\dag} \hat{H}'(\k) \hat{d}_{\k,s}, 
\end{eqnarray}
through the unitary transformation. 
 Here, $\hat{d}_{\k,s}^{\dag}=
(d_{\k,1,s}^{\dag},d_{\k,2,s}^{\dag},d_{\k,3,s}^{\dag})$ where 
$d_{\k,1,s}^{\dag}$, $d_{\k,2,s}^{\dag}$ and $d_{\k,3,s}^{\dag}$ are 
creation operators of $|a_{\rm 1g}>$, $|e_{\rm g}, 1>$ and 
$|e_{\rm g}, 2>$-orbitals, respectively.

 We choose the {\it c}-axis as a quantization axis of spin for a convenience 
of following discussion. 
 By choosing the basis functions as eqs.~(\ref{eq:a1g}-\ref{eq:e2}), 
we obtain a simplified expression for the atomic spin-orbit 
coupling term $H_{\rm LS}$ as, 
\begin{eqnarray}
&& \hspace{-10mm}  H_{\rm LS} = \lambda \sum_{\k}
\left(
\begin{array}{cc}
\hat{d}_{\k,\uparrow}^{\dag} & \hat{d}_{\k,\downarrow}^{\dag} 
\\
\end{array}
\right)
\nonumber \\
&& \hspace{-10mm} \times
\left(
\begin{array}{cccccc}
0 & 0 & 0 & 0 & {\rm i} & -1 \\
0 & 0 & -{\rm i} & -{\rm i} & 0 & 0 \\
0 & {\rm i} & 0 & 1 & 0 & 0 \\
0 & {\rm i} & 1 & 0 & 0 & 0 \\
-{\rm i} & 0 & 0 & 0 & 0 & {\rm i} \\
-1 & 0 & 0 & 0 & -{\rm i} & 0 \\
\end{array}
\right)
\left(
\begin{array}{c}
\hat{d}_{\k,\uparrow} \\
\hat{d}_{\k,\downarrow} \\
\end{array}
\right). 
\label{eq:LS-coupling}
\end{eqnarray}
 Interestingly, the matrix element of eq.~(\ref{eq:LS-coupling}) is 
the same as that in Sr$_2$RuO$_4$.~\cite{rf:ng,rf:yanaseRuSO} 
 Therefore, we can discuss the role of spin-orbit coupling in \Co 
in analogy with Sr$_2$RuO$_4$. 
 Note that the basis wave function in Sr$_2$RuO$_4$ 
is $d_{\rm xy}$-, $d_{\rm yz}$- and $d_{\rm zx}$-orbitals, 
while that is $|a_{\rm 1g}>$, $|e_{\rm g}, 1>$ and $|e_{\rm g}, 2>$-orbitals 
in \Cof. 
 This difference arises from the position of apex oxygens. 
 While an apex of RuO$_6$ octahedron is along the $c$-axis, 
all of apex oxygens in CoO$_6$ octahedron are tilted from the $c$-axis. 
 When we consider the matrix element in eq.~(\ref{eq:LS-coupling}), 
the $d_{\rm xy}$-orbital in Sr$_2$RuO$_4$ corresponds to 
the $a_{\rm 1g}$-orbital in \Cof, while  
the $d_{\rm yz}$- and $d_{\rm zx}$-orbitals 
correspond to the $e'_{\rm g}$-doublet. 
 It is expected that the $d_{\rm xy}$-orbital is responsible for the 
superconductivity in Sr$_2$RuO$_4$,~\cite{rf:nomura,rf:yanaseRuSO} 
while the superconductivity is mainly caused by the $e'_{\rm g}$-doublet 
in \Cof.~\cite{rf:yanase,rf:mochizuki} 
 Thus, these two materials provide contrasting examples of  
spin triplet superconductors.

 The coupling constant $2 \lambda$ in $H_{\rm LS}$ has been estimated as 
$57$mev~\cite{rf:michioka}. This value corresponds to 
$\lambda=0.17$ in our unit if we choose the band width 
$W \sim 9 t_{3} = 1.5$eV according to the LDA calculations. 
 Since this estimation has some ambiguities, we investigate
the range from $\lambda=0.05$ to $\lambda=0.25$. 
 In this range, the effect of spin-orbit coupling on the band structure 
is very small. 
 There is a hole pocket enclosing the $\Gamma$-point and six hole
pockets near the K-points, which are consistent with LDA 
calculations.~\cite{rf:singh,rf:pickett,rf:arita} 
 The former (\agff) mainly consists of the $a_{\rm 1g}$-orbital and the 
latter (\egff) mainly consists of the $e'_{\rm g}$-orbitals.

 From the analysis of three-orbital Hubbard model 
with $\lambda=0$,~\cite{rf:yanase,rf:mochizuki} 
we have found that the superconductivity is mainly induced by 
the $e'_{\rm g}$-orbitals and the $a_{\rm 1g}$-orbital 
can be simply ignored to discuss qualitative features of 
superconductivity. 
 Therefore, we derive a two-orbital Hubbard model by simply dropping 
the $a_{\rm 1g}$-orbital as, 
\begin{eqnarray}
\label{eq:two-orbital-model}
&& \hspace{-5mm} H_{2} = H^{(2)}_{0}+H^{(2)}_{\rm I},  
\\
&& \hspace{-5mm} H^{(2)}_{0}=
\sum_{\k,s} \hat{a}_{\k,s}^{\dag} \hat{H}^{(2)}(\k,s) \hat{a}_{\k,s}, 
\\
&& \hspace{-5mm} H^{(2)}_{\rm I} = 
 U \sum_{i} \sum_{a=1}^{2} n_{i,a,\uparrow} n_{i,a,\downarrow} 
+ U' \sum_{i} \sum_{a>b} n_{i,a} n_{i,b} 
\nonumber \\ 
&& \hspace{5mm} - 
\Jh \sum_{i} \sum_{a>b} (2 \SS_{i,a} \SS_{i,b} + \frac{1}{2} n_{i,a} n_{i,b})
\nonumber \\ 
&& \hspace{5mm}
+ J \sum_{i} \sum_{a \neq b}  
a_{i,a,\downarrow}^{\dag} 
a_{i,a,\uparrow}^{\dag} 
a_{i,b,\uparrow} 
a_{i,b,\downarrow}. 
\end{eqnarray}
 Here, we have introduced a two component vector 
$\hat{a}_{\k,s}^{\dag}=(d_{\k,2,s}^{\dag},d_{\k,3,s}^{\dag})$ and 
$2 \times 2$ matrix 
\begin{eqnarray}
\hspace{5mm} \hat{H}^{(2)}(\k,s) =
\left(
\begin{array}{cc}
e'_{11}(\k) & e'_{12}(\k) -{\rm i} s \lambda  \\
e'_{21}(\k) +{\rm i} s \lambda & e'_{22}(\k)  \\
\end{array}
\right). 
\label{eq:2by2-matrix}
\end{eqnarray}
 Here, $e'_{ij}(\k)$ is obtained from eq.~(\ref{eq:unitary-local}) 
as $e'_{ij}(\k) = \hat{H}'(\k)_{i+1,j+1}$ where $\hat{H}'(\k)_{i,j}$ 
is the matrix element of $\hat{H}'(\k)$. 
 It should be noticed that the off-diagonal matrix elements connecting 
up and down spins in eq.~(\ref{eq:LS-coupling}) vanish. 
 This means that the operators $L_{\rm x}$ and $L_{\rm y}$ have no
matrix element in the Hilbert space expanded by $e'_{\rm g}$-orbitals. 
 Therefore, only the spin-orbit coupling along the {\it z}-axis 
is effective in the two-orbital model. 
 This fact remarkably simplifies the calculation and provides a 
clear understanding for the results in \S4.

 The second term in eq.~(\ref{eq:two-orbital-model}) describes 
the on-site Coulomb interactions including the intra-orbital 
repulsion $U$, inter-orbital repulsion $U'$, Hund's rule coupling 
$\Jh$ and pair hopping term $J$. 
 We impose the relations $U=U'+\Jh+J$ and $\Jh=J$ throughout this paper. 
 The Coulomb interaction term is invariant for the unitary 
transformation with respect to the orbital owing to these relations.

 In this paper, we investigate the $d$-vector in \Co on the basis of 
the two-orbital Hubbard model in eq.~(\ref{eq:two-orbital-model}).  
 According to Refs.~37 and 38, we choose the hopping parameters as, 
\begin{eqnarray}
\label{eq:minor-matrix}
&& \hspace{-15mm} (t_1,t_2,t_4,t_5,t_6,t_7,t_8,t_9)=
\nonumber \\
&& \hspace{-5mm}
a (0.1,0.2,0.3,-0.2,-0.05,0.2,0.2,-0.25). 
\end{eqnarray}
where $a = 0.6 \sim 1$ is consistent with LDA calculations. 
 We choose a small value of $a = 0.6$ in this paper. 
 This is because the electric DOS in \egf decreases by neglecting the 
$a_{\rm 1g}$-orbital and this artificial decrease is compensated 
by decreasing $a$. 
 We define the number of holes in the two-orbital model as $n_{\rm e}$ 
which corresponds to the area of \egff. 
 According to the chemical analysis,~\cite{rf:takadanumber,
rf:karppinen} the total number of holes is $n = 0.4 \sim 0.5$ 
which is different from $n=0.67$ in the parent material 
Na$_{0.33}$CoO$_{2}$. 
 Since $0 \le n_{\rm e} \le n$, 
we investigate the range $n_{\rm e}=0.05 \sim 0.35$. 
 We have found that $n_{\rm e}$ is an important parameter for the 
superconductivity rather than $n$.~\cite{rf:yanase} 
 The superconducting instability is significantly suppressed when 
$n_{\rm e}=0$. 
 For finite $n_{\rm e}$, the superconducting \Tc slightly increases 
with the increase of crystal field splitting $e_{\rm c}$ which leads 
to the increase of $n_{\rm e}$.~\cite{rf:yanase,rf:mochizukiJPSJ} 
 The recent NMR measurement has confirmed an important role of 
$e_{\rm c}$.~\cite{rf:ihara} 
 We show a typical Fermi surface of two-orbital model in Fig.~1.

\begin{figure}[ht]
\begin{center}
\includegraphics[height=5cm]{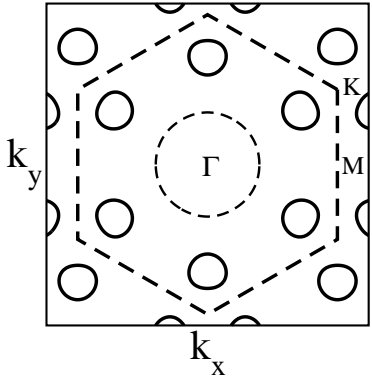}
\caption{
Fermi surface in the two-orbital Hubbard model (solid lines). 
We have shown the Fermi surface of $a_{\rm 1g}$-orbital 
which is determined as $\hat{H}'(\k)_{11}-\mu=0$ (thin dashed line). 
The thick dashed line is the first Brillouin zone of triangular 
lattice. 
The parameters are chosen to be $\lambda=0.17$, $a=0.6$, 
$n_{\rm e}=0.21$ and $n=0.5$. 
} 
\label{fig:fermisurface2by2}
\end{center}
\end{figure}

 It should be noted that ARPES measurements~\cite{rf:hasan,rf:yang} 
for non-superconducting Na$_x$CoO$_{2}$ observed the $a_{\rm 1g}$-Fermi 
surface, but the $e_{\rm g}$-Fermi surface has not been found. 
 This observation implies $n_{\rm e}=0$ which contradicts with our 
basic assumption, namely a finite value of $n_{\rm e}$. 
 Our analysis on the three-orbital Hubbard model has shown that 
the superconductivity is hardly stabilized 
when $n_{\rm e}=0$.~\cite{rf:yanase} 
 This result implies that the \egf exists in the superconducting 
materials as expected in LDA calculations. 
 The increase of $n_{\rm e}$ is actually expected in superconducting 
materials because the H$_2$O molecules increase the crystal field 
splitting $e_{\rm c}$, as shown by the 
NMR measurement.~\cite{rf:ihara}

\section{Classification of the Pairing State}

 Before the microscopic analysis, 
we classify the symmetry of pairing state in the two-orbital 
Hubbard model including the spin-orbit coupling. 
 First, we show the orbital part of spin triplet pairing function
in Table I. 
 There are two wave functions in both $p$-wave symmetry and 
$f$-wave symmetry. 
 The $p_{\rm x}$-wave and $p_{\rm y}$-wave are 
degenerate owing to the symmetry of triangular lattice. 
 On the other hand, the $f_1$-wave and $f_2$-wave are 
not degenerate. 
\begin{table}[htbp]
 \begin{center}
   \begin{tabular}{|c|c|c|} \hline 
\smash{\lower2.0ex\hbox{E$_1$}} & $p_{\rm x}$-wave & 
$\sin \frac{\sqrt{3}}{2}k_{\rm x} \cos \frac{1}{2}k_{\rm y}$
\\
\cline{2-3}
& $p_{\rm y}$-wave & 
$\sin k_{\rm y} + \sin \frac{1}{2}k_{\rm y}
\cos \frac{\sqrt{3}}{2}k_{\rm x}$
\\\hline
B$_1$ & $f_1$-wave & 
$
\sin \frac{1}{2}k_{\rm y} 
(\cos \frac{1}{2}k_{\rm y} - \cos \frac{\sqrt{3}}{2}k_{\rm x})
$
\\\hline
B$_2$ & $f_2$-wave & 
$\sin \frac{\sqrt{3}}{2}k_{\rm x} 
(\cos \frac{3}{2}k_{\rm y} - \cos \frac{\sqrt{3}}{2}k_{\rm x})
$
\\\hline
    \end{tabular}
    \caption{
Orbital part of odd-parity pairing function in the triangular lattice. 
The first column shows the irreducible representations of D$_6$-group. 
The second column shows the notations adopted in this paper, which are  
the counterparts of the isotropic system. 
The third column shows the typical wave function. Note that the 
wave function obtained in the Dyson-Gorkov equation is different from the 
third column to some extent. 
}  
  \end{center}
\end{table}

 The spin part of pairing function is described by the $d$-vector 
as,~\cite{rf:leggett,rf:sigrist}
\begin{eqnarray}
&& \hspace{-10mm} \left(
\begin{array}{cc}
\Delta_{\uparrow\uparrow}(k) & \Delta_{\uparrow\downarrow}(k) \\
\Delta_{\downarrow\uparrow}(k) & \Delta_{\downarrow\downarrow}(k) \\
\end{array}
\right)
=
{\rm i} \hat{d}(k) \hat{\sigma} \sigma_{y}
\nonumber \\
&& \hspace{-14mm}
=
\left(
\begin{array}{cc}
-d_{{\rm x}}(k)+{\rm i}d_{{\rm y}}(k) & d_{{\rm z}}(k) \\
d_{{\rm z}}(k) & d_{{\rm x}}(k)+{\rm i}d_{{\rm y}}(k) \\
\end{array}
\right). 
\end{eqnarray}
 Since the Cooper pair has spin $S=1$, there is a $3$-fold degeneracy 
in the $SU(2)$ symmetric system. 
 Therefore, if the spin-orbit coupling is neglected, the degeneracy in 
the $p$-wave state is $3 \times 2=6$-fold due to the spin part and 
orbital part, while that in the $f$-wave state is $3$-fold.

\begin{table}[htbp]
 \begin{center}
   \begin{tabular}{|c|c|} \hline 
$P_{\rm xy+}$ & 
$p_{\rm x}\hat{x}+p_{\rm y}\hat{y}$, $p_{\rm y}\hat{x}-p_{\rm x}\hat{y}$
\\\hline
$P_{\rm xy-}$ & 
$p_{\rm x}\hat{x}-p_{\rm y}\hat{y}$, $p_{\rm y}\hat{x}+p_{\rm x}\hat{y}$
\\\hline
$P_{\rm z}$ & 
$(p_{\rm x} \pm {\rm i} p_{\rm y}) \hat{z}$
\\\hline
$F_{\rm xy}$ & 
$f_1 \hat{x} - \alpha f_{2} \hat{y}$, $\alpha f_2 \hat{x} + f_{1} \hat{y}$
\\\hline
$F_{\rm z}$ & 
$f_1 \hat{z}$
\\\hline
    \end{tabular}
    \caption{
Classification of the pairing state including the spin-orbit coupling. 
The first column shows the notations adopted in this paper. 
}  
  \end{center}
\end{table}
 Table II shows the classification of pairing state for finite 
spin-orbit coupling. 
 According to the results for $\lambda=0$,~\cite{rf:yanase} 
the $f_2$-wave state is not stabilized in the multi-orbital Hubbard model. 
 Therefore, we have not shown the pairing state which is reduced to 
the $f_2$-wave state in the limit $\lambda \rightarrow 0$. 
 Since the spin-orbit coupling $\lambda < 0.25$ is much smaller than the 
band width $W \sim 9$, $|\alpha|$ in $F_{\rm xy}$-state is 
much smaller than $1$.

 The $F_{\rm xy}$-state and all of the $p$-wave states in Table II 
are two-dimensional representations. 
 Although the two pairing states in $P_{\rm xy+}$ are  
not degenerate in general,~\cite{rf:sigrist} there remains an additional 
degeneracy in the two-orbital Hubbard model. 
 This is because the matrix element of $L_{\rm x}$ and $L_{\rm y}$ 
vanishes in this model. 
 The symmetry of Hubbard Hamiltonian with $\lambda=0$ is 
$G \otimes SU(2) \otimes T \otimes U(1)$, where $G$, $SU(2)$, $T$ and 
$U(1)$ show the symmetries of point group, spin rotation, time-reversal 
and gauge transformation, respectively. 
 When the spin-orbit coupling exists, the symmetry is reduced to 
$G \otimes U(1) \otimes T \otimes U(1)$. 
 Thus, the $SU(2)$ symmetry is violated by the spin-orbit coupling.  
 However, the $U(1)$ symmetry for the spin rotation in the plane 
remains because $L_{\rm x}=L_{\rm y}=0$. This additional $U(1)$ symmetry 
is the origin of degeneracy in $P_{\rm xy+}$-state. 
 In the three-orbital Hubbard model
(eq.~(\ref{eq:multi-orbital-model})), 
the symmetry is reduced to $G \otimes T \otimes U(1)$, 
and the degeneracy in the $P_{\rm xy+}$-state is lifted. 
 However, it is expected that the lifting of degeneracy is small 
because the order parameter in $a_{\rm 1g}$-orbital is very 
small.~\cite{rf:mochizuki,rf:yanase}

\section{Determination of the Pairing State}

\subsection{Linearized Dyson-Gorkov equation without $SU(2)$ symmetry}

 In this section, we determine the pairing state in the 
two-orbital Hubbard model 
by solving the linearized Dyson-Gorkov equation within the second order 
perturbation theory. 
 The derivation of Dyson-Gorkov equation has been explained in 
literatures.~\cite{rf:yanasereview} 
 The application to the multi-orbital model including the 
spin-orbit coupling has been given in the study of 
Sr$_2$RuO$_4$.~\cite{rf:yanaseRuSO} 
 The application to the two-orbital model 
(eq.~(\ref{eq:two-orbital-model})) is simpler than that to 
Sr$_2$RuO$_4$ because the transverse component of spin-orbit 
coupling vanishes.

 In order to make following discussions clear, we introduce a 
unitary matrix $\hat{U}(\k,s)=(u_{ij}(\k,s))$ which 
diagonalizes $\hat{H}^{(2)}(\k,s)$, namely, 
\begin{eqnarray}
 \label{eq:unitary}
\hat{U}^{\dag}(\k,s)  \hat{H}^{(2)}(\k,s) \hat{U}(\k,s) 
= 
\left(
\begin{array}{cc}
E_1(\k) & 0 \\
0 & E_2(\k) \\
\end{array}
\right).
\end{eqnarray}
 Here, $E_1(\k) \leq E_2(\k)$ and $E_{i}(\k)$ do not depend on 
the spin owing to the time-reversal symmetry and inversion symmetry. 
 The Green function $\hat{G}(k,s) = 
({\rm i}\omega_{n} \hat{1} - \hat{H}^{(2)}(\k,s))^{-1}$ 
is obtained as, 
\begin{eqnarray}
 \label{eq:Green-function}
G_{ij}(k,s)=\sum_{\alpha=1}^{2} u_{i\alpha}(\k,s) u^{*}_{j\alpha}(\k,s) 
G_{\alpha}(k),
\end{eqnarray}
where $G_{\alpha}(k)=\frac{1}{{\rm i}\omega_{n}-
E_{\alpha}(\mbox{{\scriptsize \boldmath$k$}})}$.

\begin{figure}[ht]
\begin{center}
\includegraphics[height=3.8cm]{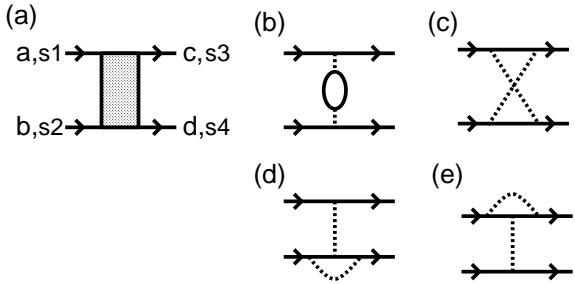}
\caption{
(a) Diagrammatic representation of the irreducible vertex in the 
particle-particle channel. 
(b-e) The second order terms with respect to the Coulomb interactions 
(dashed lines). The solid line denotes the Green function having the 
indices of spin and orbital. 
} 
\label{fig:diagram}
\end{center}
\end{figure}

 As shown in the LDA calculations, the energy band described by 
$E_2(\k)$ crosses the Fermi level and $E_1(\k)$ is far below 
the Fermi level. 
 In this case, the superconducting transition is induced by 
the Cooper pairing in the $E_2(\k)$-band. 
 Therefore, the linearized Dyson-Gorkov equation is written in terms of an  
effective interaction within the $E_2(\k)$-band, 
\begin{eqnarray}
&& \hspace{-10mm}
\lambda_{\rm e} \Delta_{s1 s2}(k) = 
- \sum_{k',s3,s4} V(k,k',s1,s2,s3,s4) 
\nonumber \\
&& \hspace{24mm}
\times
|G_{2}(k')|^{2} \Delta_{s3 s4}(k'), 
\label{eq:eliashberg-equation} 
\end{eqnarray}
with 
\begin{eqnarray}
\label{eq:effective-interaction}
&& \hspace{-15mm}
V(k,k',s1,s2,s3,s4)=
\sum_{abcd} u^{*}_{a2}(\k,s1) u^{*}_{b2}(-\k,s2) 
\nonumber \\
&& \hspace{-12mm}
\times 
V_{abcd}(k,k',s1,s2,s3,s4) u_{c2}(\k',s3) u_{d2}(-\k',s4). 
\end{eqnarray}
 Here, $V_{abcd}(k,k',s1,s2,s3,s4)$ is the irreducible vertex in the 
particle-particle channel having the indices of orbital and spin 
(see Fig.~2(a)). 
 Because of $L_{\rm x}=L_{\rm y}=0$, the $z$-component of spin is 
conserved and $V(k,k',s1,s2,s3,s4)$ is finite only if $s1+s2=s3+s4$. 
 Therefore, the linearized Dyson-Gorkov equation is reduced to the eigenvalue 
equations for 
$\Delta_{\uparrow \uparrow}$, $\Delta_{\uparrow \downarrow}$
and $\Delta_{\downarrow \downarrow}$, respectively. 
\begin{eqnarray}
\label{eq:eliashberg-equation-uu} 
&&  \hspace{-10mm}
\lambda^{\uparrow \uparrow}_{\rm e} \Delta_{\uparrow \uparrow}(k) = 
- \sum_{k'} V_{\uparrow \uparrow}(k,k') |G_{2}(k')|^{2} 
\Delta_{\uparrow \uparrow}(k'), 
\\
\label{eq:eliashberg-equation-ud} 
&& \hspace{-10mm}
\lambda^{\uparrow \downarrow}_{\rm e} \Delta_{\uparrow \downarrow}(k) = 
- \sum_{k'} V_{\uparrow \downarrow}(k,k') |G_{2}(k')|^{2} 
\Delta_{\uparrow \downarrow}(k'), 
\\
&& \hspace{-10mm}
\lambda^{\downarrow \downarrow}_{\rm e} \Delta_{\downarrow \downarrow}(k) = 
- \sum_{k'} V_{\downarrow \downarrow}(k,k') |G_{2}(k')|^{2} 
\Delta_{\downarrow \downarrow}(k'), 
\label{eq:eliashberg-equation-dd} 
\end{eqnarray}
where $V_{\uparrow \uparrow}(k,k')=
V(k,k',\uparrow,\uparrow,\uparrow,\uparrow)$, 
$V_{\downarrow \downarrow}(k,k')=
V(k,k',\downarrow,\downarrow,\downarrow,\downarrow)$ 
and $V_{\uparrow \downarrow}(k,k')=
V(k,k',\uparrow,\downarrow,\uparrow,\downarrow)+
V(k,k',\uparrow,\downarrow,\downarrow,\uparrow)$.  
 It is notable that the maximum eigenvalues of 
eqs.~(\ref{eq:eliashberg-equation-uu}) and (\ref{eq:eliashberg-equation-dd}) 
are equivalent. 
 The wave functions are related to be $\Delta_{\downarrow \downarrow}(k) 
= e^{{\rm i}\phi} \Delta^{*}_{\uparrow \uparrow}(k)$ where 
the phase $\phi$ is arbitrary.

 If we neglect the spin-orbit coupling, the maximum eigenvalue 
does not depend on the direction of $d$-vector 
since $V_{\uparrow \uparrow}(k,k')=V_{\downarrow \downarrow}(k,k')=
V_{\uparrow \downarrow}(k,k')$. 
 However, we find $\lambda^{\uparrow \uparrow}_{\rm e} = 
\lambda^{\downarrow \downarrow}_{\rm e} \ne 
\lambda^{\uparrow \downarrow}_{\rm e}$ when $\lambda$ is finite. 
 The superconducting transition is determined by the criterion,  
max$(\lambda^{\uparrow \uparrow}_{\rm e},
\lambda^{\uparrow \downarrow}_{\rm e}) = 1$.  
 If $\lambda^{\uparrow \uparrow}_{\rm e} >
\lambda^{\uparrow \downarrow}_{\rm e}$, the $P_{\rm xy+}$-, 
$P_{\rm xy-}$- or $F_{\rm xy}$-state is stabilized 
at $T=T_{\rm c}$, while $P_{\rm z}$- or $F_{\rm z}$-state 
is stabilized when $\lambda^{\uparrow \uparrow}_{\rm e} < 
\lambda^{\uparrow \downarrow}_{\rm e}$.

 In this paper, we estimate irreducible vertex 
$V_{abcd}(k,k',s1,s2,s3,s4)$ up to the second order with respect to 
the Coulomb interaction term $H^{(2)}_{\rm I}$. 
 The diagrammatic representation is shown in Figs.~2(b-e) which are 
calculated from the possible combination of Coulomb interactions 
and Green functions. 
 We numerically solve the eigenvalue equations for 
$\lambda^{\uparrow \uparrow}_{\rm e}$ and 
$\lambda^{\uparrow \downarrow}_{\rm e}$ and determine the pairing state 
with highest \Tcf. 
 In the following, we divide the first Brillouin zone into 
$128 \times 128$ lattice points and we take 512 Matsubara frequencies 
for $T \geq 0.005$ and 1024 Matsubara frequencies for $0.002 \leq T < 0.005$. 
  We have confirmed that this size of calculation is sufficient for 
the following results.

\subsection{$d$-vector}

 First, we show the phase diagram of two-orbital Hubbard model 
without spin-orbit coupling which will be a basis of 
following discussions. 
 Figure~3 shows the stable pairing symmetry in the phase diagram 
of $n_{\rm e}$ and $\Jh$. 
 In the figure, the intra-orbital repulsion $U$ is determined 
so as to obtain $T_{\rm c}=0.01$ which 
corresponds to $20$K in our unit. 
 This temperature is higher than the observed 
transition temperature $T_{\rm c} \sim 5$K. 
 However, the stable pairing symmetry is almost 
independent of the temperature as shown in Ref.~38.

\begin{figure}[htbp]
\begin{center}
\includegraphics[width=7.2cm]{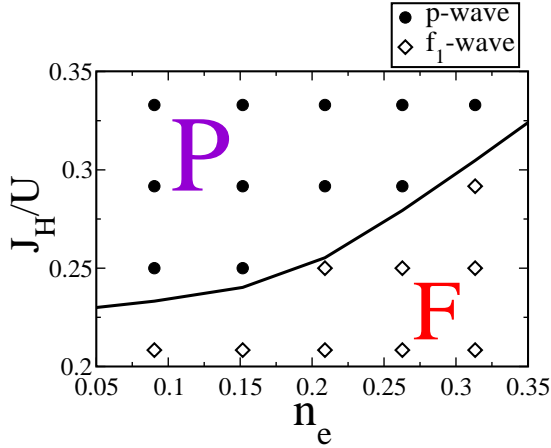}
\caption{Phase diagram without spin-orbit coupling 
in the $n_{\rm e}$-$\Jh$ plane. 
The solid line is the phase boundary obtained by the interpolation. 
} 
\label{fig:lamdazero}
\end{center}
\end{figure}

 Among many parameters in the multi-orbital Hubbard model, 
the relevant parameter for the pairing symmetry is 
the Hund's rule coupling $\Jh$ and the number of holes in 
\egf $n_{\rm e}$.~\cite{rf:yanase} 
 Figure~3 is quite similar to the phase diagram of three-orbital model 
shown in Ref.~38. 
 When the Hund's rule coupling is large and/or $n_{\rm e}$ is small, 
the $p$-wave superconductivity is stabilized. 
 The $f$-wave superconductivity is stabilized for small $\Jh$ and/or
large $n_{\rm e}$. 
 The only qualitative difference between the two-orbital and 
three-orbital models is that the $f$-wave state stabilized for 
$n_{\rm e} < 0.1$ and $\Jh/U > 0.2$ in Ref.~38 vanishes in Fig.~3.

\begin{figure}[htbp]
\begin{center}
\includegraphics[width=8.5cm]{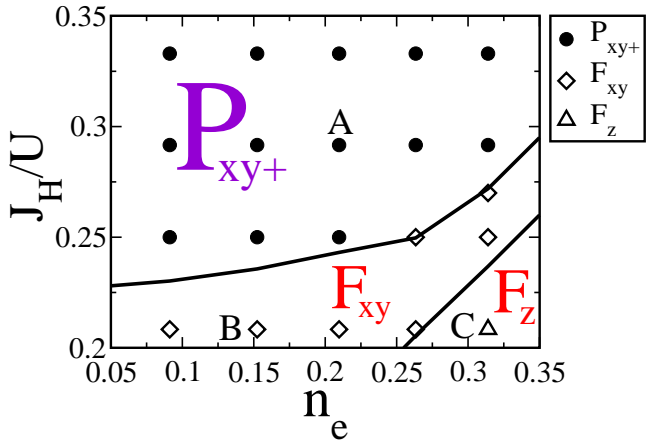}
\caption{
Phase diagram including the spin-orbit coupling $\lambda=0.17$. 
The notations of $P_{\rm xy+}$, $F_{\rm xy}$ and $F_{\rm z}$ have been 
given in Table II. 
The solid line is the phase boundary obtained by the interpolation. 
} 
\label{fig:a06}
\end{center}
\end{figure}

 We show the results of $d$-vector in Fig.~4, where the \Tc is fixed to 
be $T_{\rm c} =0.003 \sim 6$K consistent with experimental value. 
 Hereafter, the parameter $U$ is determined so as to obtain this \Tcf. 
 For example, we obtain $U=6.57$, $U=5.47$ and $U=4.34$ in A, B and C in
Fig.~4, respectively. 
 It is clearly shown that the spin-orbit coupling stabilizes 
the $P_{\rm xy+}$-state when the pairing symmetry is $p$-wave. 
 Then, the $d$-vector is parallel to the plane. 
 On the other hand, the $d$-vector in the $f$-wave symmetry 
depends on the parameters, although the $F_{\rm xy}$-state seems to be 
most stable.

 In order to  understand these results, it is useful to analyze the 
role of spin-orbit coupling in a perturbative way. 
 Although we have included the spin-orbit coupling term into the 
unperturbed Hamiltonian, the perturbative treatment is 
valid since $\lambda$ is much smaller than the band width. 
 As a result of perturbation expansion in $\lambda$, 
the effective interaction $V_{s s'}(k,k')$ is written as, 
\begin{eqnarray}
\label{eq:first-order} 
V_{s s'}(k,k')=V^{(0)}(k,k') 
+ \sum_{n=1}^{\infty} \lambda^{n} V^{(n)}_{s s'}(k,k'). 
\end{eqnarray}

 We have actually applied this perturbation theory to Sr$_2$RuO$_4$ 
and shown that the first order term vanishes when the $\gamma$-band 
is mainly superconducting.~\cite{rf:yanaseRuSO} 
 This is because the hybridization term disappears between 
$d_{\rm xy}$-orbital and $d_{\rm yz}$-  $d_{\rm zx}$-orbitals. 
 However, the first order term exists in case of \Co since the 
large hybridization term exists in the $e'_{\rm g}$-doublet.

 It is easy to find that the first order term always appears with 
the combination to the off-diagonal Green function 
or off-diagonal matrix element of $\hat{U}(\k,s)$. 
 Taking into account the symmetry of $e'_{12}(\k)$ in 
eq.~(\ref{eq:2by2-matrix}), namely $e'_{12}(k_{\rm x},k_{\rm y})=
-e'_{12}(-k_{\rm x},k_{\rm y})=-e'_{12}(k_{\rm x},-k_{\rm y})$, 
we obtain the following relations, 
\begin{eqnarray}
\label{eq:relation1} 
&& \hspace{-10mm}
V^{(1)}_{\uparrow \uparrow}(k,k')=
V^{(1)*}_{\downarrow \downarrow}(k,k'), 
\\
\label{eq:relation2} 
&& \hspace{-10mm}
V^{(1)}_{\uparrow \uparrow}(k_{\rm x},k_{\rm y},k'_{\rm x},k'_{\rm y})
=-V^{(1)}_{\uparrow \uparrow}(-k_{\rm x},k_{\rm y},-k'_{\rm x},k'_{\rm y})
\nonumber \\
&& \hspace{19mm}
=-V^{(1)}_{\uparrow \uparrow}(k_{\rm x},-k_{\rm y},k'_{\rm x},-k'_{\rm y}), 
\\
\label{eq:relation3} 
&&  \hspace{-10mm} 
V^{(1)}_{\uparrow \downarrow}(k,k')=0. 
\end{eqnarray}
 According to eq.~(\ref{eq:relation2}), the kernel of 
linearized Dyson-Gorkov equation works on the $p_{\rm x}$-wave state 
to produce the $p_{\rm y}$-wave state, namely 
\begin{eqnarray}
\label{eq:px-py-coupling}
\sum_{k,k'} \Delta_{\rm y}(k)
V^{(1)}_{\uparrow \uparrow}(k,k') |G_{2}(k')|^{2} 
\Delta_{\rm x}(k') \ne 0,
\end{eqnarray} 
and
\begin{eqnarray}
\label{eq:px-px-coupling}
&& \hspace{-15mm} 
\sum_{k,k'} \Delta_{\rm x}(k)
V^{(1)}_{\uparrow \uparrow}(k,k') |G_{2}(k')|^{2} 
\Delta_{\rm x}(k') = 
\nonumber \\
&& \hspace{-15mm}
\sum_{k,k'} \Delta_{\rm y}(k)
V^{(1)}_{\uparrow \uparrow}(k,k') |G_{2}(k')|^{2} 
\Delta_{\rm y}(k') = 0,
\end{eqnarray} 
where $\Delta_{\rm x}(k)$ and $\Delta_{\rm y}(k)$ are functions 
having the symmetry of $p_{\rm x}$-wave and $p_{\rm y}$-wave, respectively.  
 Thus, the first order term in $\lambda$ works like a first order 
perturbation on the degenerate states, namely  $p_{\rm x}$- and 
$p_{\rm y}$-states. Therefore, the eigenvalue 
$\lambda_{\rm e}^{\uparrow \uparrow}$ for one of the $P_{\rm xy+}$- 
and $P_{\rm xy-}$-states increases in the first order of $\lambda$, 
while $\lambda_{\rm e}^{\uparrow \uparrow}$ 
for the other state decreases. 
 On the other hand, the first order term vanishes for the $d$-vector
along the {\it c}-axis according to eq.~(\ref{eq:relation3}). 
 Therefore, the $d$-vector along the plane is always stabilized within 
the first order theory in $\lambda$ independent of the microscopic 
details. 
 Which state is stabilized between $P_{\rm xy+}$ and $P_{\rm xy-}$
is determined by the microscopic calculation for $V_{s s'}(k,k')$. 
 Figure~4 shows that the $P_{\rm xy+}$-state is stable within 
the second order perturbation theory.

 The situation is quite different for the $f$-wave superconductivity. 
 Owing to eq.~(\ref{eq:relation2}), 
the first order term is not effective in the $f_1$-wave state, namely 
\begin{eqnarray}
\label{eq:f1-f1-coupling}
\sum_{k,k'} \Delta_{\rm f1}(k)
V^{(1)}_{\uparrow \uparrow}(k,k') |G_{2}(k')|^{2} 
\Delta_{\rm f1}(k') = 0, 
\end{eqnarray} 
where $\Delta_{\rm f1}(k)$ is the function having the $f_1$-wave 
symmetry. 
 Although the first order term couples the $f_1$-wave state 
to the $f_2$-wave state, this is a higher order effect because 
these states are non-degenerate. 
 Indeed, the eigenvalue of linearized Dyson-Gorkov equation 
for the $f_2$-wave state is much smaller than that for the $f_1$-wave 
state.~\cite{rf:yanase} 
 Then, the role of first order term, 
$V^{(1)}_{\uparrow \uparrow}(k,k')$ is negligibly small. 
 Therefore, the direction of $d$-vector is determined by 
the higher order terms beyond the second order,  
namely $V^{(n)}_{s s'}(k,k')$ ($n \ge 2$). 
 The role of higher order terms depends on the microscopic details 
as we have shown in the case of Sr$_2$RuO$_4$.~\cite{rf:yanaseRuSO} 
 In \Cof,  $F_{\rm xy}$-state ($F_{\rm z}$-state) is stabilized for 
small (large) value of $n_{\rm e}$ and/or large (small) value of 
$\Jh$ as shown in Fig.~4.

\begin{figure}[htbp]
\begin{center}
\includegraphics[width=7.5cm]{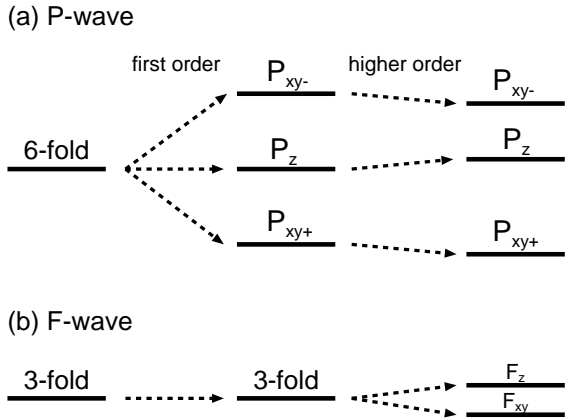}
\caption{
Schematic figure of level scheme for (a) $p$-wave superconductivity 
and (b) $f$-wave superconductivity. The degeneracy at $\lambda=0$ 
is shown on the left. The level splitting due to the first order 
perturbation in $\lambda$ is shown in the center. 
The energy levels affected by the higher order perturbation are 
shown on the right. We assume in (b) that the $F_{\rm xy}$-state is 
stable rather than the $F_{\rm z}$-state. 
} 
\label{fig:tcdiff}
\end{center}
\end{figure}

 We summarize the above discussions in Fig.~5. 
 In case of the $p$-wave superconductivity, the first order perturbation 
in $\lambda$ stabilizes the $P_{\rm xy+}$-state and destabilizes the 
$P_{\rm xy-}$-state. The $P_{\rm z}$-state is not affected in this 
order. We find that higher order perturbation terms favor 
the $P_{\rm xy+}$-state furthermore, as shown in Fig.~5(a). 
 In case of the $f$-wave superconductivity, three-fold degeneracy 
is not lifted by the first order perturbation. 
 The higher order perturbation stabilizes the $F_{\rm xy}$-state 
or $F_{\rm z}$-state depending on the parameters such as $\Jh$ 
and $n_{\rm e}$.

\subsection{Splitting of \Tc}

 The quite different role of spin-orbit coupling between the $p$- and
$f$-wave symmetries discussed above is illuminated by showing the 
splitting of \Tcf. 
 In 
eqs.~(\ref{eq:eliashberg-equation-uu}-\ref{eq:eliashberg-equation-dd}), 
the linearized Dyson-Gorkov equation is defined for each direction of 
$d$-vector. 
 We define the superconducting \Tc for 
$\hat{d}\parallel${\it xy} and that for $\hat{d}\parallel${\it z} 
by the criterion  
$\lambda_{\rm e}^{\uparrow \uparrow}(T_{\rm c}^{\rm (xy)})=1$
and 
$\lambda_{\rm e}^{\uparrow \downarrow}(T_{\rm c}^{\rm (z)})=1$, 
respectively. 
 The former corresponds to the \Tc of $P_{\rm xy+}$- or 
$F_{\rm xy}$-state while the latter is the \Tc of $P_{\rm z}$- or 
$F_{\rm z}$-state.

 Figure~6 shows the splitting of \Tcf, namely $\Delta T_{\rm c}/T_{\rm c} 
=|T_{\rm c}^{\rm (xy)}-T_{\rm c}^{\rm (z)}|/T_{\rm c}$. 
 The parameters $\Jh$ and $n_{\rm e}$ are chosen to be A, B, C in
Fig.~4, where the $P_{\rm xy+}$-state, $F_{\rm xy}$-state and 
$F_{\rm z}$-state are stable, respectively. 
 It is clearly shown that $\Delta T_{\rm c}$ is much smaller in the 
$f$-wave symmetry than in the $p$-wave symmetry. 
 This is because the first order term in $\lambda$ is ineffective. 
 Indeed, while $\Delta T_{\rm c}/T_{\rm c}$ increases linearly in 
the $p$-wave symmetry, the increase in the $f$-wave symmetry is 
nearly proportional to the square of $\lambda$.

\begin{figure}[htbp]
\begin{center}
\includegraphics[width=7.5cm]{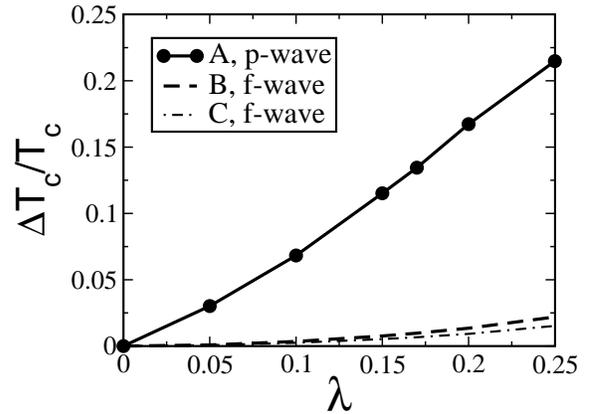}
\caption{
$\lambda$-dependence of $\Delta T_{\rm c}/T_{\rm c}
=|T_{\rm c}^{\rm (xy)}-T_{\rm c}^{\rm (z)}|/T_{\rm c}$. 
The parameters are chosen to be A, B and C in Fig.~4. 
The $P_{\rm xy+}$-state, $F_{\rm xy}$-state and $F_{\rm z}$-state
are stable, respectively.  
} 
\label{fig:tcdiff}
\end{center}
\end{figure}

 The quite different magnitude of $\Delta T_{\rm c}/T_{\rm c}$ shown 
in Fig.~6 will be reflected in the phase diagram under the magnetic field. 
 In order to gain the Zeeman coupling energy, the $d$-vector 
can be rotated by the applied magnetic field. 
 Fig.~6 indicates that the $d$-vector is strongly fixed against the 
magnetic field in case of the $p$-wave symmetry, 
however the $d$-vector in the $f$-wave symmetry rotates 
in a weak magnetic field. 
 We have actually determined the phase diagram under the magnetic field 
on the basis of the weak coupling theory and found that the rotation 
does not occur in the $p$-wave symmetry up to the half of Pauli 
paramagnetic limit.~\cite{rf:yanaseII} 
 The results for the multiple phase diagram under the 
magnetic field and the characteristics of each phase will be shown 
in the subsequent publication.~\cite{rf:yanaseII}

 The results of Figs.~4 and 6 are important for an interpretation 
of NMR and $\mu$SR results because the magnetic susceptibility has 
an anisotropy arising from the direction of 
$d$-vector.~\cite{rf:leggett,rf:sigrist} 
 In general, the magnetic susceptibility in the unitary state is 
obtained as, 
\begin{eqnarray}
\label{eq:susceptibility}
&& \hspace*{-12mm}
\chi_{\mu \nu}/\chi_{\rm n} = 
\nonumber \\
&& \hspace*{-12mm}
<\delta_{\mu \nu} - 
(d_{\mu}(\k) d_{\nu}(\k)/|\hat{d}(\k)|^{2})(1-Y(\k,T))>_{\rm F}, 
\end{eqnarray} 
where $<>_{\rm F}$ is an average on the Fermi surface, 
$<A>_{\rm F} = \rho(0)^{-1}\int A \delta(E_{2}(\k)) {\rm d}k$. 
Here, $\chi_{\rm n}$ is the magnetic susceptibility in the normal state, 
and $Y(\k,T)$ is the angle dependent Yosida function,
\begin{eqnarray}
\label{eq:yoshida}
Y(\k,T)=\int_{-\infty}^{\infty} {\rm d} \xi 
(-\frac{{\rm d}f(E)}{{\rm d}E})|_{E=\sqrt{\xi^{2}+|\hat{d}(k)|^{2}}}, 
\end{eqnarray} 
where $f(E)$ is the Fermi distribution function. 
 It should be noticed that the magnetic susceptibility decreases 
for the magnetic field parallel to the $d$-vector. 
 For the $P_{\rm xy+}$-state, it is reasonable to assume 
that $\hat{d}=p_{\rm x} \hat{x} + p_{\rm y} \hat{y}$ 
or $\hat{d}=p_{\rm y} \hat{x} - p_{\rm x} \hat{y}$ is stabilized 
below \Tc among any linear combinations of these states. 
 This is because the condensation energy is maximally 
gained in these states within the weak coupling theory. 
 Then, the susceptibility along the plane 
$\chi_{\rm ab}$ decreases to the half of its value in the normal state, 
as shown in Fig.~7. 
 In order to obtain Fig.~7, we have assumed the order parameter below 
\Tc as 
$\hat{d}(\k)=\Delta(T)(\phi_{\rm x}(\k) \hat{x} + \phi_{\rm y}(\k) \hat{y})$
where $\phi_{\rm x}(\k)$ and $\phi_{\rm y}(\k)$ are obtained 
in eq.~(\ref{eq:eliashberg-equation-uu}) as 
$\Delta_{\uparrow \uparrow}(\k,{\rm i}\pi T_{\rm c}) 
= - \phi_{\rm x}(\k) + {\rm i} \phi_{\rm y}(\k)$. 
 The wave functions $\phi_{\rm x}(\k)$ and $\phi_{\rm y}(\k)$ have the 
symmetry of $p_{\rm x}$-wave and $p_{\rm y}$-wave, respectively. 
 We have calculated the temperature dependence of $\Delta(T)$ 
by using the effective model having the separable pairing interaction and 
solving it within the BCS theory. 
 The same procedure has been used in Ref.~59. 
 For the $F_{\rm xy}$-state, the in-plane magnetic field 
favors one of the degenerate states. For example, 
$\hat{d}=\alpha f_{2} \hat{x}+f_{1}\hat{y}$ is favored by the magnetic field 
along {\it x}-axis. Then, $\chi_{\rm ab}$ decreases owing to 
the induced $f_{2}$-wave component, however the decrease is very small 
as shown in Fig.~7 because $|\alpha| \ll 1$. 
 Here, we have calculated the temperature dependence of order parameter 
in the same way as for the $P_{\rm xy+}$-state. 
 The susceptibility along {\it c}-axis $\chi_{\rm c}$ 
does not decrease in both $P_{\rm xy+}$- and $F_{\rm xy}$-states. 
 Although $\chi_{\rm c}$ decreases to zero in the $F_{\rm z}$-state at 
$T=0$, the observation of this decrease will be difficult because 
the $d$-vector rotates in a very weak magnetic field.

\begin{figure}[htbp]
\begin{center}
\includegraphics[width=7cm]{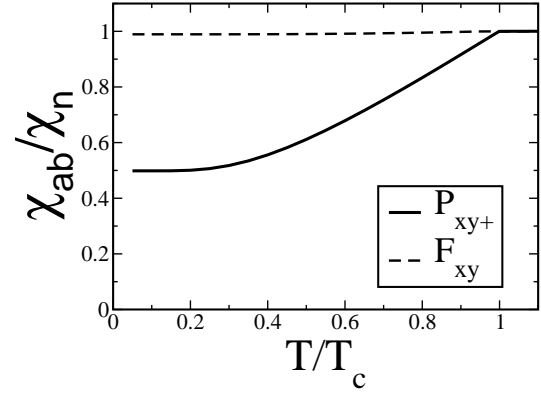}
\caption{Temperature dependence of magnetic susceptibility 
along the plane $\chi_{\rm ab}$ divided by the normal state value 
$\chi_{\rm n}$. We consider the magnetic field along {\it x}-axis and 
assume $\hat{d}=p_{\rm x} \hat{x} + p_{\rm y} \hat{y}$ 
in the $P_{\rm xy+}$-state and 
$\hat{d}=\alpha f_{2} \hat{x}+f_{1}\hat{y}$ in the $F_{\rm xy}$-state. 
 We obtain the Fermi surface and momentum dependence of order parameter 
for the parameters A and B in Fig.~4, respectively. 
} 
\label{fig:kaiab}
\end{center}
\end{figure}

 The Knight shift below \Tc has been measured by 
some groups.~\cite{rf:yoshimura,rf:kobayashi,rf:ishidaprivate,
rf:zhengprivate,rf:higemoto}
 Most of the measurements have been performed under the magnetic field  
parallel to the plane. 
 It has been reported that the Knight shift in Co-NMR and O-NMR 
decreases below \Tc in a weak magnetic field.~\cite{rf:kobayashi,
rf:yoshimura,rf:ishidaprivate,rf:zhengprivate} 
 This observation is consistent with the $P_{\rm xy+}$-state in 
our results. The $P_{\rm xy+}$-state is furthermore consistent with 
the $\mu$SR measurement~\cite{rf:higemoto} which does not detect any 
time-reversal symmetry breaking expected in the $P_{\rm z}$-state 
and $d_{\rm x^{2}-y^{2}} + {\rm i} d_{\rm xy}$-state. 
 Contrary to the other NMR measurements, the Knight shift in D-NMR is 
almost temperature independent below \Tcf.~\cite{rf:yoshimura} 
 Although this observation is incompatible with the other NMR measurements, 
this is consistent with $F_{\rm xy}$- and $F_{\rm z}$-states in our 
calculation.

\subsection{Crossover from $\lambda \ll W$ to $\lambda \gg W$}

 In \S4.1-3, we have considered the spin-orbit coupling 
much smaller than the band width, namely $\lambda \ll W$. 
 This relation is expected in most of the $d$-electron systems 
including \Cof. 
 On the other hand, the situation is opposite in the $f$-electron 
systems which include odd-parity superconductors. 
 It is not clear whether the role of spin-orbit coupling is 
qualitatively different between the weak coupling region 
$\lambda \ll W$ and strong coupling region $\lambda \gg W$. 
 Therefore, it is interesting to investigate global behaviors 
of two-orbital Hubbard model 
from $\lambda \ll W$ to $\lambda \gg W$, although the strong coupling 
region $\lambda \gg W$ is unrealistic for \Cof.

 In Fig.~8, we show the anisotropy of eigenvalues of 
linearized Dyson-Gorkov equation $\lambda_{\rm an} = 
\lambda_{\rm e}^{\uparrow \downarrow}/\lambda_{\rm e}^{\uparrow \uparrow}$.
 The small value of $\lambda_{\rm an}$ means that the splitting of 
superconducting \Tc is large. 
 It is shown that the splitting of \Tc takes 
its maximum for an intermediate value of $\lambda$. 
 Thus, the anisotropy of $d$-vector decreases with increasing 
the spin-orbit coupling in the strong coupling region $\lambda \gg W$
in contrast to the weak coupling region. 
 This behavior is quite different from the 
anisotropy of spin susceptibility 
$\chi_{\rm an}=\chi_{\rm c}/\chi_{\rm ab}$ which has been shown 
in Fig.~8 for a comparison. 
 The anisotropy of spin susceptibility is enhanced monotonically 
with increasing the spin-orbit coupling, as expected.~\cite{rf:comment} 
 Note that the structure around $\lambda=2$ is due to the rapid change 
of Fermi surface. 
 There are two hole pocket Fermi surfaces enclosing the 
K-point for $\lambda > 2$.

\begin{figure}[htbp]
\begin{center}
\includegraphics[width=7.5cm]{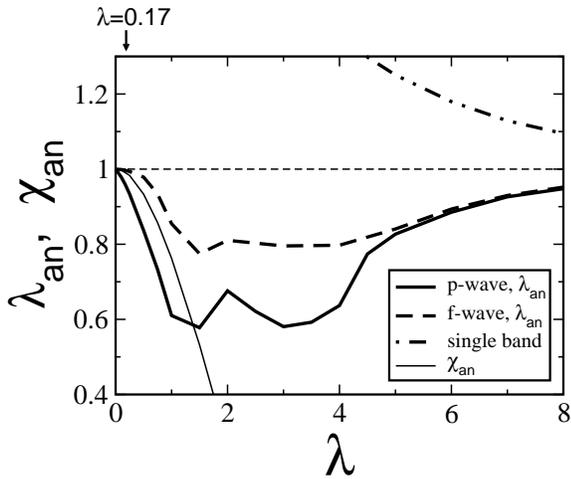}
\caption{The anisotropy of eigenvalues of linearized Dyson-Gorkov 
equation, 
$\lambda_{\rm an} = 
\lambda_{\rm e}^{\uparrow \downarrow}/\lambda_{\rm e}^{\uparrow \uparrow}$ 
in a wide range of spin-orbit coupling. 
 The solid (dashed) line shows the result for the $p$-wave ($f$-wave) 
symmetry. The parameters are 
$\Jh/U=7/24$, $e_{\rm c}=0$, ($\Jh/U=5/24$, $e_{\rm c}=-0.5$) 
and $T=0.01$. Here, we fix the crystal field splitting 
$e_{\rm c}$ and total hole density $n=0.5$ instead of $n_{\rm e}$. 
 The anisotropy of spin susceptibility 
$\chi_{\rm an}=\chi_{\rm ab}/\chi_{\rm c}$ is shown by the  
thin solid line. 
 The double dotted line shows 
$\lambda_{\rm e}^{\rm s}/\lambda_{\rm e}^{\uparrow \uparrow}$ 
where $\lambda_{\rm e}^{\rm s}$ is the eigenvalue in the 
single-orbital Hubbard model (eq.~(\ref{eq:single-orbital-model})). 
} 
\label{fig:grobal}
\end{center}
\end{figure}

 It is easy to understand the behavior of $\lambda_{\rm an}$ 
by taking the limit $\lambda \rightarrow \infty$ in advance. 
 In this limit, the two-orbital Hubbard model is reduced to 
the single-orbital Hubbard model written as, 
\begin{eqnarray}
H_{1} =\sum_{\k,s} E_{2}(\k) c_{\k,s}^{\dag} c_{\k,s} + 
\frac{U+U'}{2} \sum_{i} n_{i,\uparrow} n_{i,\downarrow}, 
\label{eq:single-orbital-model}
\end{eqnarray}
where the Coulomb interaction is renormalized to $\frac{U+U'}{2}$. 
 It is notable that $s$ in eq.~(\ref{eq:single-orbital-model}) is not 
the spin but the pseudo-spin. 
 This single-orbital model has the $SU(2)$ symmetry for the rotation of 
pseudo-spin. Therefore, spin triplet superconducting state is isotropic 
for the direction of $d$-vector defined by the pseudo-spin. 
 In Fig.~8, we see that $\lambda_{\rm an}$ in the two-orbital model 
comes to unity in the limit $\lambda \rightarrow \infty$. 
 Then, the maximum eigenvalue of linearized Dyson-Gorkov equation 
actually comes to that of single-orbital Hubbard model. 
 On the other hand, the $g$-factor of Zeeman coupling term is 
anisotropic for the pseudo-spin, and therefore the anisotropy of 
spin susceptibility remains in the single-orbital model. 
 In other words, the effect of spin-orbit coupling is absorbed 
by the character of quasi-particles and the residual interaction 
is isotropic in the strong coupling limit 
$\lambda \rightarrow \infty$.

 Although the two-orbital Hubbard model in 
eq.~(\ref{eq:two-orbital-model}) is too simple to provide 
a general formula, the same situation is generally expected 
when the heavy quasi-particles consist of one Kramers doublet. 
 Actually, it has been reported by the NMR measurements~\cite{rf:tou} 
that the $d$-vector in UPt$_3$ is almost isotropic, although the 
spin susceptibility is significantly anisotropic. 
 Thus, the anisotropy of $d$-vector and that of spin susceptibility 
is quite different in general.

\section{Comparison with Sr$_2$RuO$_4$}

 Interestingly, we have found some similar aspects 
between \Co and Sr$_2$RuO$_4$. The latter is the most established 
spin triplet superconductor in $d$-electron systems.~\cite{rf:maeno} 
 One of the similarity is the orbital dependent superconductivity. 
 The three \tg is divided to two groups, namely 
$d_{\rm xy}$-orbital and $d_{\rm yz}$- $d_{\rm zx}$-orbitals in 
Sr$_2$RuO$_4$, $a_{\rm 1g}$-orbital and $e'_{\rm g}$-orbitals in 
\Cof. 
 The superconductivity is mainly induced by the former in Sr$_2$RuO$_4$ 
while by the latter in \Cof. 
 Another interesting finding is that the spin-orbit coupling term 
has the same matrix element as eq.~(\ref{eq:LS-coupling}). 
 This enables us to summarize the results on the $d$-vector in a 
unified way as shown in Table~III.

\begin{table}[htbp]
 \begin{center}
   \begin{tabular}{|c|c|c|c|} \hline 
\multicolumn{2}{|c|}{Sr$_2$RuO$_4$} & \multicolumn{2}{|c|}{\Cof}
\\\hline
\multicolumn{2}{|c|}{Square lattice} & \multicolumn{2}{|c|}{Triangular lattice}
\\\hline
$d_{\rm xy}$ & $d_{\rm yz}$, $d_{\rm zx}$ & 
\multicolumn{2}{|c|}{$e'_{\rm g}$}
\\\hline
$L_{\rm x}$, $L_{\rm y}$ & $L_{\rm z}$ & \multicolumn{2}{|c|}{$L_{\rm z}$}
\\\hline
\multicolumn{2}{|c|}{$p$-wave} & $p$-wave & $f$-wave
\\\hline
O($\lambda^{2}$) & O($\lambda$) & O($\lambda$) & O($\lambda^{2}$)
\\\hline
$\hat{d}\parallel$z & $\hat{d}\parallel$xy & $\hat{d}\parallel$xy 
& both 
\\\hline
    \end{tabular}
    \caption{
An unified description on the $d$-vector between Sr$_2$RuO$_4$ and 
\Cof. The second column shows the symmetry of crystal. 
The third column shows the orbitals leading to the superconductivity. 
The forth column shows the important component of orbital moment. 
The fifth column shows the symmetry of superconductivity. 
The sixth column shows the leading order term with respect to the 
spin-orbit coupling. 
The last column shows the direction of $d$-vector. 
The results on Sr$_2$RuO$_4$ have been obtained in Ref.~50. 
}  
  \end{center}
\end{table}

 We also show the case where the $d_{\rm yz}$- and 
$d_{\rm zx}$-orbitals induce the superconductivity in Sr$_2$RuO$_4$. 
 In this case, the results on the $d$-vector are qualitatively the same 
as the $p$-wave superconductivity in \Cof. 
 Actually, the analysis of the first order term in $\lambda$ (see \S4.2) 
can be applied to the case in Sr$_2$RuO$_4$ in the same 
way.~\cite{rf:yanaseRuSO} 
 In both cases, the orbital moment along the {\it c}-axis stabilizes 
the $d$-vector along the plane.

 The first order term in $\lambda$ vanishes in Sr$_2$RuO$_4$ when 
the $d_{\rm xy}$-orbital is active. Although this is similar to 
the $f$-wave state in \Cof, the microscopic origin is quite different.  
 The first order term disappears in Sr$_2$RuO$_4$ since the hybridization 
term with $d_{\rm yz}$- and $d_{\rm zx}$-orbitals 
vanishes.~\cite{rf:yanaseRuSO}

 Note that the results on Sr$_2$RuO$_4$ are consistent with experiments. 
 It is believed that the $d_{\rm xy}$-orbital is mainly superconducting
in Sr$_2$RuO$_4$. 
 Then, we obtain the chiral superconducting state,~\cite{rf:yanaseRuSO}
namely $P_{\rm z}$-state in Table II, which is consistent with 
the $\mu$SR measurement.~\cite{rf:luke} 
 Owing to the disappearance of the first order term in $\lambda$, 
the $d$-vector can be rotated by a weak magnetic field along 
the {\it c}-axis. 
 Actually, the NMR measurement has detected a $d$-vector along 
the plane under the weak magnetic field $H\parallel c$.~\cite{rf:murakawa}

\section{Summary and Discussions}

 In this paper, we have investigated the $d$-vector in the 
possible spin triplet superconductor \Co on the basis of the 
two-orbital Hubbard model representing the $e'_{\rm g}$-orbitals. 
 There remains a $6$-fold ($3$-fold) degeneracy in the $p$-wave 
($f$-wave) superconducting state if we neglect the spin-orbit coupling. 
 Therefore, we include the L-S coupling term in Co ions into the 
Hamiltonian and determine the pairing state on the basis of 
the second order perturbation theory. 

 We find that the role of spin-orbit coupling is quite different 
between the $p$-wave superconductivity and $f$-wave superconductivity. 
 The $d$-vector is always along the plane if the orbital part has the 
$p$-wave symmetry. 
 On the other hand, the direction of $d$-vector 
depends on the parameters if the orbital part has 
the $f$-wave symmetry. 
 In our calculation, the superconducting \Tc is determined for each 
direction of $d$-vector. 
 We show that the splitting of \Tc in the $p$-wave state is much larger 
than that in the $f$-wave state. 

 Such a different role of spin-orbit coupling is explained 
by analyzing the first order term with respect to the 
spin-orbit coupling $\lambda$. This term is effective in 
the $p$-wave symmetry, but ineffective in the $f$-wave symmetry. 
 This property is hold in all of the perturbation terms with respect to 
the Coulomb interactions. 
 Therefore, the results in this paper will be valid beyond the second 
order perturbation theory adopted here. 
 The only assumption is that the superconductivity is mainly induced 
by the $e'_{\rm g}$-orbitals.

 The determination of $d$-vector is especially important 
for the interpretation of Knight shift measurements 
because the magnetic properties of spin triplet superconductor 
depend on the direction of $d$-vector. 
 Our results indicate that the $d$-vector can be rotated by 
a weak magnetic field in the case of $f$-wave superconductivity, 
while the $d$-vector is strongly fixed against the applied field in 
the case of $p$-wave superconductivity. 
 The NMR Knight shift along the plane will decrease 
in the latter case because the $d$-vector has both $\hat{x}$ and 
$\hat{y}$ components. 
 On the other hand, the Knight shift will be almost 
temperature-independent in the former case in all directions of 
applied magnetic field. 
 Unfortunately, experimental results seem to be confusing. 
 The NMR Knight shift in Co-site ~\cite{rf:yoshimura,rf:kobayashi,
rf:ishidaprivate,rf:zhengprivate} and O-site ~\cite{rf:ishidaprivate} 
has shown a sizable decrease for the parallel magnetic field.
 This is consistent with our results for the $p$-wave state. 
 On the other hand, the Knight shift in D-site has reported a 
qualitatively different result~\cite{rf:yoshimura} which is consistent 
with our results for the $f$-wave state.

 Finally, we suggest some future experimental studies 
which are highly desired. 
 First, the NMR Knight shift along {\it c}-axis 
may provide conclusive evidence for the pairing symmetry. 
 Kobayashi {\it et al.} has reported a pioneering result which shows 
a slight decrease of Knight shift below \Tcf.~\cite{rf:kobayashi} 
 Although this observation implies the spin singlet superconducting 
state, the theoretical interpretation seems not to be conclusive. 
 It is noted that the decrease of {\it c}-axis Knight shift  
in Ref.~7 is very small compared with that of in-plane Knight shift. 
 Although this small decrease may be explained on the basis of the 
vortex state with spin singlet pairing,~\cite{rf:kobayashi} 
another theoretical interpretation may be also valid. 
 For example, we suggest three possibilities based on the spin triplet 
superconducting state. 
 The first one is the coexistence of $p$- and $f$-wave 
superconductivities which will be discussed in the subsequent 
paper.~\cite{rf:yanaseII} Extending the present theory to determine the 
pairing state below \Tcf, we find this coexistent state around the 
boundary between the $p$- and $f$-wave states in Fig.~4. 
 The magnetic susceptibility decreases in this state for all 
directions of magnetic field. 
 The second one is the role of strong electron correlation. 
 The conclusions obtained in the weak coupling BCS theory can be 
modified in the strongly correlated electron systems, especially near
the magnetic instability. 
 The third one is the role of disorder which leads to the rotation of 
$d$-vector. 
 The detailed measurements of {\it c}-axis Knight shift including 
the magnetic field dependence, sample dependence and impurity effect 
are highly desired for a clear identification of the pairing state.

 Second, the search for the multiple phase transition under 
the magnetic field is particularly interesting in the future study. 
 The phase transition accompanied by a rotation of $d$-vector is 
expected in the spin triplet superconductor under the magnetic field. 
 Such a phase transition is promising when the $H_{\rm c2}$ is large, 
namely when the magnetic field is applied along the plane. 
 We have actually determined the phase diagram 
under the parallel magnetic field on the basis of the weak coupling 
theory.~\cite{rf:yanaseII} We find that such a phase transition is expected 
in the $p$-wave superconducting state and $p+f$ coexistent state 
while that is not expected in the $f$-wave state. 
 Interestingly, the second phase in the $p$-wave state is different from 
all of the pairing states in Table~II, and the NMR Knight shift 
decreases even in this phase. The results on this subject will 
be reported in another publication.~\cite{rf:yanaseII}

 Third, the precise measurement on the critical field $H_{\rm c2}$ 
along the plane is important. 
 Although some authors have reported the measurement, 
the results are controversial.~\cite{rf:sasaki,rf:sakuraireview}
 If the $H_{\rm c2}$ far exceeds the Pauli paramagnetic limit, 
the spin triplet pairing state is promising. 
 This issue is closely related to the second issue, namely the 
possibility of multiple phase transition. 
 For example, it has been reported that the Knight shift in the Co-NMR 
and $\mu$SR does not decrease in a high magnetic field along the 
plane.~\cite{rf:yoshimura,rf:higemoto} 
 If this magnetic field is still below $H_{\rm c2}$, the qualitatively 
different behavior from the low magnetic field region indicates 
a multiple phase transition.

 At last, it is desirable to determine the Fermi surface of  
superconducting material experimentally. 
 Although the existence of \egf is a basic assumption of this paper, 
it is not clear whether the \egf exists or not. 
 Recent ARPES measurements~\cite{rf:hasan,rf:yang} on the 
non-superconducting material have reported a qualitatively 
different Fermi surface from the LDA calculations.~\cite{rf:singh,
rf:pickett,rf:arita} 
 We think that the measurement on the superconducting material 
is highly desired.

\section*{Acknowledgments}

 The authors are grateful to Y. Ihara, H. Ikeda, K. Ishida, M. Kato, 
Y. Kitaoka, Y. Kobayashi, C. Michioka, K. Miyake, Y. Ono, 
N. E. Phillips, H. Sakurai, M. Sato, M. Udagawa, Y. J. Uemura, 
K. Yamada, K. Yada, K. Yoshimura and G-q. Zheng 
for fruitful discussions. 
 Numerical computation in this work was partly carried out 
at the Yukawa Institute Computer Facility. 
 The present work has been partly supported by a Grant-In-Aid for 
Scientific Research from the Ministry of Education, Culture, Sport, 
Science and Technology of Japan.

\end{document}